# RECONFIGURABLE MID-INFRARED HYPERBOLIC METASURFACES USING PHASE-CHANGE MATERIALS

T. G. Folland[1], A. Fali[2], S. T. White[3], J. R. Matson[4], S. Liu[5], N. A. Aghamiri[2], J. H, Edgar[5], R. F. Haglund, Jr.[3,4], Y. Abate[2*] and J. D. Caldwell[1+]

[1] Department of Mechanical Engineering, Vanderbilt University, Nashville, TN 37212
[2] Department of Physics and Astronomy, University of Georgia, Athens, GA 30602-2451
[3] Department of Physics and Astronomy, Vanderbilt University, Nashville, TN 37235
[4] Interdisciplinary Materials Science Program, Vanderbilt University, Nashville, TN 37212
[5] Department of Chemical Engineering, Kansas State University, Manhattan, KS 66506
*yabate@physast.uga.edu  +josh.caldwell@vanderbilt.edu

**Metasurfaces offer the potential to control light propagation at the nanoscale for applications in both free-space and surface-confined geometries[1-3]. Existing metasurfaces frequently utilize metallic polaritonic[4] elements with high absorption losses[5], and/or fixed geometrical designs that serve a single function. Here we overcome these limitations by demonstrating a reconfigurable hyperbolic[6] metasurface comprising of a heterostructure of isotopically enriched hexagonal boron nitride (hBN)[7-10] in direct contact with the phase-change material (PCM) vanadium dioxide ($VO_2$)[11-14]. Spatially localized metallic and dielectric domains in $VO_2$ change the wavelength of the hyperbolic phonon polaritons (HPhPs) supported in hBN by a factor 1.6 at 1450cm$^{-1}$. This induces in-plane launching, refraction and reflection of HPhPs in the hBN, proving reconfigurable control of in-plane HPhP propagation at the nanoscale[15]. These results exemplify a generalizable framework based on combining hyperbolic media and PCMs in order to design optical functionalities such as resonant cavities, beam steering, waveguiding and focusing with nanometric control.**

Engineering reconfigurability into optical materials significantly increases the versatility of devices and the range of their potential applications[1,2]. Phase change materials (PCMs) are an appealing approach to doing so as they undergo significant changes in optical properties upon exposure to external stimuli[16]. By integrating PCMs and polaritonic materials, changes in optical properties induced by a phase transition can control polariton dispersion, which can in turn be exploited in reconfigurable metasurfaces[16-20]. However, one of the phases in PCMs is typically metallic and/or exhibits high optical losses. In previous studies of surface-confined polaritons, polariton propagation was restricted to spatial regions where a low-loss dielectric phase was present[16,17]. Here we significantly reduce PCM-induced losses by using isotopically enriched[9,10] hexagonal boron nitride (hBN), a natural hyperbolic medium that supports HPhPs.

Hyperbolic polaritons remain sensitive to local changes in the dielectric function of the ambient environment[21], but the electromagnetic near-fields are strongly confined to the volume of the hyperbolic material[6-8]. This means that HPhPs interact with spatially localized phase transitions of a PCM, yet do not suffer significant optical losses from this interaction, and thus can be supported over both metallic and dielectric phases. Through electromagnetic modelling, we demonstrate that such PCM-HPhP heterostructures can be designed as optical resonators[8,22] and metasurfaces[23,24], as well as refractive near-field components, such as waveguides and lenses. Most significantly, for reversible PCM transitions, any design can be fully reconfigured.

The prototype device (Fig. 1a) consists of a 24 nm thick flake of $^{10}$B-enriched hBN (~99% enriched [9,10]) transferred using low-contamination transfer techniques onto a single crystal of $VO_2$ grown on quartz. We use scattering-type scanning near-field optical microscopy (s-SNOM) to directly map and visualize evanescent optical fields, corresponding to polaritonic waves of compressed wavelength $\lambda_p$, propagating within the hBN (see Fig. 1a). In s-SNOM images, HPhPs can be observed in two ways: first, polaritons launched by the tip reflect from boundaries in the sample (e.g. a flake edge) creating interference fringes with spacing $\lambda_p/2$, which are scattered back to free-space by the tip and detected[7,25,26]. Second, polaritons can be directly launched from sample edges and propagate across the surface to interfere with the tip, producing fringes with spacing $\lambda_p$[9,27]. The s-SNOM maps contain a superposition of both so-called 'tip-launched' and 'edge-launched' fringes, and are interpreted by considering the fringe spacing from individual waves ($\lambda_p/2$ vs $\lambda_p$) and the direction of polariton propagation.

In Fig. 1b we show a near-field amplitude s-SNOM image of the propagating polariton in hBN on top of the $VO_2$ crystal using a 1450cm$^{-1}$ excitation laser. This plot exhibits both a tip-launched mode propagating away from the edge of the hBN flake with wavelength $\lambda_p/2$ (purple, in the *x* direction of Fig. 1b), and an edge-launched mode propagating away from the $VO_2$ crystal edge with wavelength $\lambda_p$ (light blue, the *y* direction of Fig. 1b). This combination of tip- and edge-launched waves arise from two properties of the sample. First, the small size (440nm thickness, 6.5μm width) of the $VO_2$ crystal provides sufficient momentum to scatter robustly into polariton modes in hBN at the crystal edges[9,27]. Second, the interface between $VO_2$ and air at the crystal edge will reflect less than 100% of the polariton in hBN, which suppresses tip-launched waves, as observed in prior work[17,21].

Propagation of HPhPs is strongly influenced by the local dielectric environment[21,23], so we investigated the influence of the $VO_2$ phase transition by measuring the s-SNOM response of the sample as a function of temperature, traversing the full dielectric-to-metal transition between 60-80°C[14]. Individual $VO_2$ domains are directly observable with s-SNOM due to the dielectric contrast between domains, with metallic (dielectric) $VO_2$ appearing as bright (dark) regions (Fig 1c)[11-14]. As the device is heated further (Fig. 1d), the hBN-supported HPhPs propagate over both the metallic and dielectric domains of $VO_2$, for appreciable propagation distances in both regions. This contrasts with earlier devices based on PCMs, where polaritons propagated for only a few cycles over the dielectric phase, but were precluded from propagation over the metallic regions[17]. We attribute this difference in propagation between present and prior work to the volume-confinement of the local electromagnetic near-fields within the low-loss hBN[8], which prevents polaritonic fields from being absorbed by the lossy metallic phase of $VO_2$. Note the metal-insulator phase-domain boundaries also appear to act as polariton reflectors and launching sites, despite no measurable change in out-of-plane topography (see Supplementary Fig. S1). The large permittivity difference between metallic and insulating phases of $VO_2$ therefore presents an excellent platform to manipulate and control polariton propagation within hyperbolic materials.

When the s-SNOM maps the evanescent fields of propagating HPhP waves in the presence of multiple interfaces, complex images result from the superposition of waves launched and reflected by each domain boundary, crystal edge and the s-SNOM tip. The simplest polaritons to identify are the modes launched from the edge of the $VO_2$ crystal, as they form straight fringes aligned parallel with crystal edge, with a different polariton wavelength $\lambda_p$ above each domain. This mode launched over the dielectric (metallic) domain is shown by light blue (red) arrow in Figs. 1 c and d. These images show that the HPhP wavelength is tuned from $\lambda/12.9$ to $\lambda/20.4$ by the PCM at 1450cm$^{-1}$, the first report of the dispersion in a hyperbolic material being tuned by a PCM. Furthermore, in Fig. 1c-d, s-SNOM images show that HPhPs are directly launched where the hBN is situated over dielectric (orange arrows) and metallic (green arrows) domains, despite there being no appreciable change in the topography of the $VO_2$ crystal (Fig. S1). Whilst past work has shown that PCMs can launch polaritons[17], here they are launched over both phases, with a different wavelength over each, promising the potential for reconfigurably tuning the HPhP properties.

This heterostructure also enables the transmission of polaritons across the aforementioned domain boundaries. To simplify the analysis, domain geometries with only a single boundary are needed. As the positions of domain boundaries induced via thermal methods in $VO_2$ are naturally semi-random, we implemented multiple heating and cooling cycles to produce single-domain structures for study, examples shown in Figs. 2a,b (see also Fig. S2). Such 'reconfiguring' of the metasurface has been repeated more than eight times in our experiments, with no appreciable change in the dielectric properties of either of the two phases of $VO_2$ or the hBN flake, demonstrating the repeatability of this process.

Of particular interest is the polariton wave front that propagates away from the $VO_2$ crystal edge in the y-direction (purple dashed line): it meets the domain boundary and distorts, propagating in a direction that is not normal either to the domain or crystal edge. This is a signature of planar polariton refraction as the wave changes direction due to the local change in dielectric environment. Whilst planar polariton refraction has been reported previously for plasmon polaritons[28], this is the first direct observation of refraction for hyperbolic polaritons, and the first to study the refraction as a function of incident (transmitted) polariton angle.

If a hyperbolic polariton traverses the boundary between different $VO_2$ domains, the angle of propagation changes to conserve momentum in accordance with Snell's law[29]:

$$\frac{\sin(\theta_I)}{\sin(\theta_R)} = \frac{n_2}{n_1} \qquad (1)$$

where $n_1$ and $n_2$ are the wave effective indices in the first and second media, and $\theta_I$ and $\theta_R$ are the corresponding angles of incidence and refraction. To demonstrate that the experimentally measured images are due to refraction, we compare the results in Fig. 2b to a simplified electromagnetic simulation (Fig. 2c). Here HPhPs excited at the edge of the $VO_2$ crystal (blue) propagate in the y-direction within the dielectric phase. When these HPhPs approach the angled dielectric-metallic domain boundary (black line) some of the wave will be reflected (brown) and some will be transmitted across the boundary (black) and refracted due to the mismatch in wavevectors for the HPhPs supported over the two PCM domains. The simulation also shows waves launched directly from the domain boundary (orange and green) in Figs. 1c and d. The refracted wave will not propagate normal to either the edge of the crystal or the domain boundary – but have the same wavelength as the wave launched in the hBN by scattering of incident light off the metallic $VO_2$ crystal edge. This is indeed what is shown in our experiments by the corresponding line profiles in Fig. 2 d-f. However, the wave reflected by the metal-dielectric domain boundary is not observed here due to interference with the edge-launched mode shown in light blue. Despite this, the good agreement between Fig 2b and c shows clear evidence of HPhP refraction.

To quantify the change in the polariton wavevector and HPhP refraction induced by the $VO_2$ domains and to demonstrate the ability to reconfigure the metasurface, we systematically studied the wavelength dependence on incident frequency and refracted angle in different domain geometries. In the first case, we systematically record s-SNOM images in both metallic and dielectric domains as the monochromatic pump-laser frequency is varied and subsequently extract the wavelength through Fourier analysis (see Fig. S3) of s-SNOM images, as has been reported previously [7,9,25,26]. The experimentally extracted polariton wavevector (symbols) agrees well with numerical calculations of the HPhP dispersion for thin hBN slabs on a substrate (Fig. 3a and b)[7,9] . The dramatic change in wavevector between domains at the same incident frequency is attributable to the large change in dielectric constant in $VO_2$ between the two PCM states, which further compresses the polariton wavelength. From the measured change in polariton wavelength, calculated the ratio of the indices of refraction, $n_2/n_1$, to determine the expected angle of refraction for the HPhP waves from Eq. (1), and compared to the refracted angle extracted from the s-SNOM images in Fig. 2 and S2, to test the adherence to Snell's law for HPhPs (Fig 3c). This result is consistent with numerical simulations at a range of different angles and frequencies (see Fig. S4) confirming that Snell's law holds for HPhPs propagating across domain boundaries.

Such systematic investigation of polariton propagation and refraction at multiple angles was not possible in prior work[28] and thus, the results presented here demonstrate that the tools and concepts of refractive optics are applicable in near-field optical design as well. Indeed, the repeatable nature of both the change in polariton wavelength and Snell's law demonstrates that this platform can steer polariton propagation by proper design of the underlying local dielectric environment.

The ability to control HPhPs propagating across phase-domain boundaries opens several possibilities for engineering lithography-free metasurfaces and near-field optics. As an example motivated by prior work[17], we investigate the possibility for creating nano-resonators using this technique, where a periodic array of metallic square domains is created inside the $VO_2$ crystal underneath the hBN (in Fig. 4a.) In the simulated spectra, we observe peaks corresponding to a series of HPhP modes that can be tuned by changing the width (Fig. 4a) and periodicity (not shown) of the metallic domain. Thus, in principle, by controlling the size and shape of the metallic domain, one can realize a resonant response previously only observed in nanofabricated structures of hBN[8,30].

Refraction of HPhPs across boundaries also opens the toolbox for near-field optics to include those of conventional refractive systems, for example in-plane lenses – whereby polaritons are focused to a point via refraction. A simulation of such a lens is shown in Fig. 4b, where HPhPs are launched at the left crystal edge and propagate inward to a hemispherical metallic domain, after which they are focused to a spot. Here the combination of hyperbolic media and PCMs is critical, because for conventional surface polaritons, the high losses of the PCM metallic state would preclude polariton propagation and thus the polariton refraction required to induce focusing. While experimentally we demonstrate the principle of this reconfigurable nano-optics platform using heterostructures comprising thin slabs of hBN on $VO_2$ single crystals, this approach can readily be generalized to other materials. To demonstrate this, we have simulated a nanophotonic waveguide using both VO$_2$ and GeSbTe[17] as the underlying PCMs (See Fig. S5). The non-volatile nature of the phase change in GeSbTe[17], where both states of the PCM are stable at room temperature, offers significant benefits for laser-writing-based approaches aimed at realizing complicated nanophotonic architectures.

In conclusion, we have for the first time experimentally demonstrated that the dispersion of HPhPs can be controlled using the permittivity changes inherent in the different phases of PCMs. This enables the direct launching, reflection, transmission and refraction of HPhP waves at the domain boundaries between the phases of the PCM, due to the large change in HPhP wavelength (here a factor x1.6) that occurs for modes propagating in the hBN over each of these domains. By thermally cycling of the hBN-VO$_2$ heterostructure creates a range of PCM domain boundary geometries, enabling the demonstration of various near-field phenomena. By inducing well-defined domain structures, it is possible to design reconfigurable HPhP resonators and refractive optics in a planar format at deep subdiffraction-limited dimensions. Beyond the implications for integrated nanophotonics, reconfigurable HPhP resonators could be used to match resonant frequencies to local molecular vibrational modes for the realization of dynamic surface-enhanced infrared absorption (SEIRA)[31] spectroscopy. Ultimately, we anticipate that the combination of low-loss, hyperbolic materials and latchable PCMs could see applications in lithography-free design and fabrication of optical devices.

## METHODS

### Device Fabrication

Vanadium dioxide (VO$_2$) single crystals were grown by physical vapor transport in a quartz tube furnace at 810°C under 1.7 Torr Ar gas at a flow rate of 25 sccm. Vanadium pentoxide (V$_2$O$_5$) powder (~0.3g, Sigma Aldrich 221899) was placed in a quartz boat (10 x 1 x 1 cm) upstream of substrates and heated for 1 hr. Evaporated V$_2$O$_5$ was reduced to VO$_2$ in this process and deposited on quartz (0001) substrates. Representative crystals from each sample were investigated using Raman spectroscopy to identify the VO$_2$ phase and optical microscopy to verify the thermal phase transition. Smaller, loose crystals located on the substrate surface were removed by adhesion to a heated (60°C) layer of PMMA firmly brought into contact with the sample and subsequently retracted.

The isotopically enriched hBN crystals were grown from high-purity elemental B$^{10}$ (99.22 at%) powder by using the metal-flux method. A Ni-Cr-B powder mixture at respective 48 wt%, 48 wt%, and 4 wt% was loaded into an alumina crucible and placed in a single-zone furnace. The furnace was evacuated and then filled with $N_2$ and forming gas (5% hydrogen in balance argon) to a constant pressure of 850 Torr. During the reaction process, the $N_2$ and forming gases continuously flowed through the system with rates of 125 sccm and 25 sccm, respectively. All the nitrogen in the hBN crystal originated from the flowing $N_2$ gas. The forming gas was used to minimize oxygen and carbon impurities in the hBN crystal. After a dwell time of 24 hours at 1550 °C, the hBN crystals were precipitated onto the metal surface by cooling at a rate of 1 °C/h to 1500 °C, and then the system was quickly quenched to room temperature. Bulk crystals were exfoliated from the metal surface using thermal release tape. Crystals were subsequently mechanically exfoliated onto a PMMA/PMGI polymer bilayer on silicon. Flakes were then transferred from the polymer substrate onto $VO_2$ single crystals using a semi-dry technique, and the polymer membrane was removed using acetone and isopropyl alcohol.

## Numerical Simulations

Numerical simulations were conducted in CST Studio Suite 2017 using the frequency domain solver using plane waves incident at 45 degrees and Floquet boundary conditions. In these simulations, polariton modes were only launched by scattering from edges in the simulation, and field profiles were extracted using frequency monitors. All results used thicknesses consistent with that measured in topographic maps of the samples. Dielectric functions were taken from Ref. [9] for isotopically enriched hBN, from Ref. [14,32] for $VO_2$ and from [33] for GeSbTe.

## s-SNOM Measurements

Near-field nano-imaging experiments were carried out in a commercial (www.neaspec.com) scattering-scanning near-field microscope (s-SNOM) based around a tapping-mode atomic-force microscope (AFM). A metal-coated Si-tip of apex radius R ≈ 20 nm that oscillates at a frequency of Ω ≈ 280 kHz and tapping amplitude of about 100 nm is illuminated by monochromatic QCL laser beam at a wavelength λ=6.9 μm and at an angle $45^o$ to the sample surface. Scattered light launches hBN HPhPs in the device, and the tip then re-scatters light (described more completely in the main text) for detection in the far-field. Background signals are efficiently suppressed by demodulating the detector signal at the second harmonic of the tip oscillation frequency and employing pseudo-heterodyne interferometric detection.

## AUTHOR CONTRIBUTIONS
R.F.H, Y.A. and J.D.C. conceived and guided the experiments. S.T.W. grew the $VO_2$ crystals and identified the phase domains. S. L. and J.H.E grew the hBN crystals. T.G.F and S.T.W. fabricated the hBN $VO_2$ heterostructure. A.F. and N.A. performed s-SNOM mapping experiments of the sample at various temperatures and incident frequencies. T.G.F. advised on experimental questions, developed the electromagnetic models and analyzed s-SNOM data to show the presence of refraction. T.G.F also conducted electromagnetic simulations of resonators, lenses and waveguides.  S.T.W and J.R.M. analyzed s-SNOM data and calculated the dispersion curves.  All authors contributed to writing the manuscript.

## ACKNOWLEDGEMENTS

The authors thank Misha Fogler for providing a script to calculate the dispersion of HPhPs. T.G.F. and S.T.W. thank the staff of the Vanderbilt Institute for Nanoscience (VINSE) for technical support during fabrication, and Kiril Bolotin for design of a 2D transfer tool. Support for the $B^{10}$-enriched hBN crystal growth was provided by the National Science Foundation, grant number CMMI 1538127. YA and NA gratefully acknowledge support provided by the Air Force Office of Scientific Research (AFOSR) grant number FA9559-16-1- 0172. The work of A.F. is supported by the National Science Foundation grant 1553251.

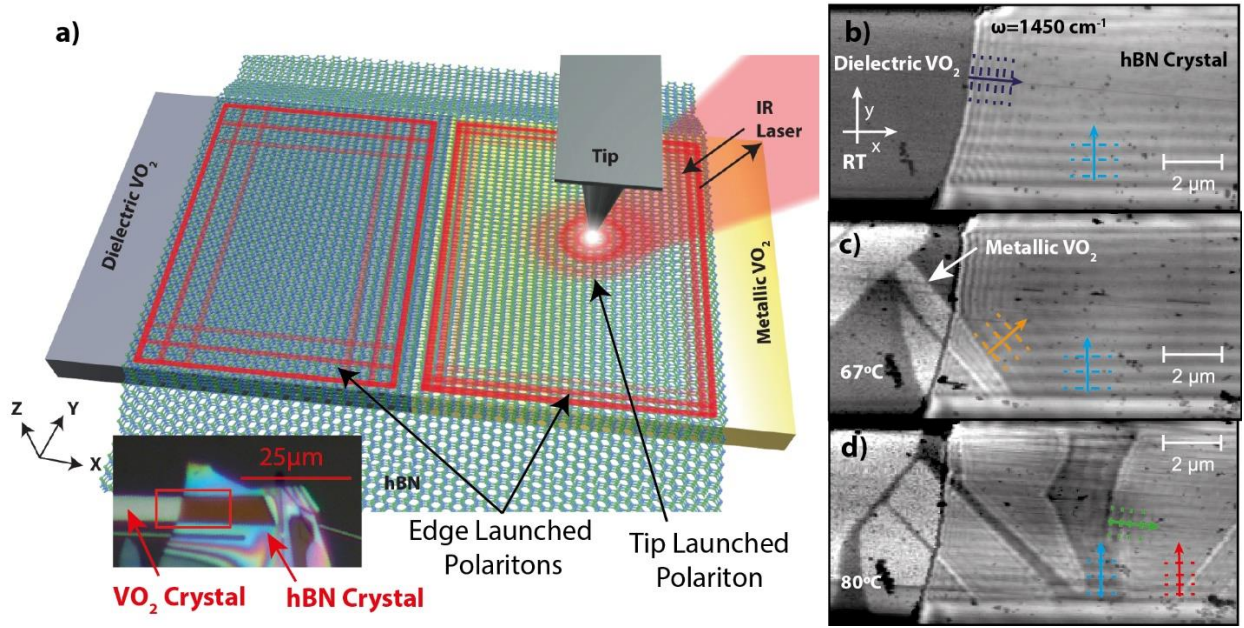

Figure 1. Actively reconfigurable hyperbolic metasurface device. (a) shows a device and experimental schematic, in which hBN has been transferred on top of a VO$_2$ single crystal and polaritons are imaged by the s-SNOM tip (b-d) s-SNOM images of the optical near-field at 1450cm$^{-1}$ (6.9 μm) at various temperatures, showing HPhPs propagating over both metallic and dielectric VO$_2$ domains. The complex patterns that form are the consequence of multiple interference from the different domains. The arrows show the following: purple highlights tip-launched modes reflected from the hBN edge, while blue designates the HPhP propagating interior to the hBN from the edge of the VO$_2$ crystal (boundary with air, suspended hBN). The red highlights the same propagation characteristics as the blue arrow, except for HPhPs propagating over the metallic VO$_2$ domains. Finally, the orange and green arrows designate HPhPs propagating within the hBN from the domain boundaries between the dielectric and metallic domains of the VO$_2$, with the orange (green) propagating over the dielectric (metallic) domains.

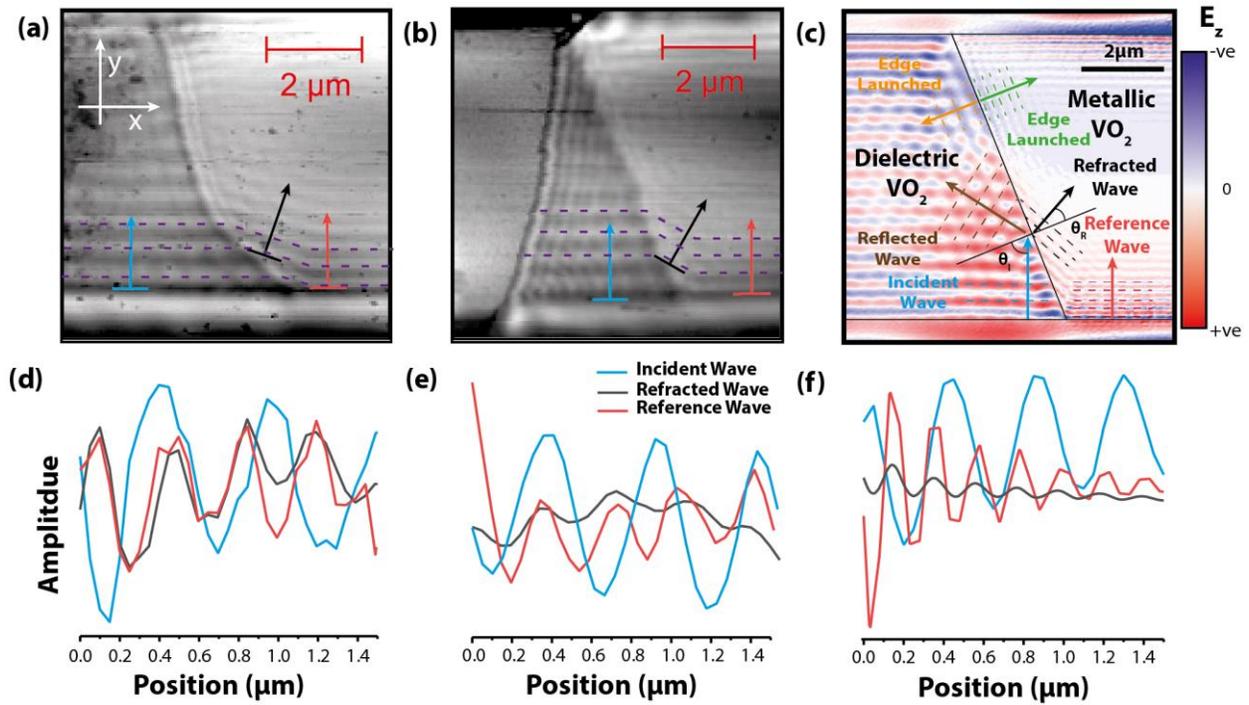

Figure 1. Hyperbolic polariton refraction on a hBN-VO$_2$ heterostructure. (a)-(b) show two s-SNOM maps of the near-field amplitude in the region of the domain boundary showing refraction. Purple dashes show the distorted phase front that propagates over the boundary. (c) shows an electromagnetic-field simulation of the geometry in (b), showing reflected, refracted and edge-launched waves. (d)-(f) show line profiles from (a)-(c) respectively, showing refraction of the wave.

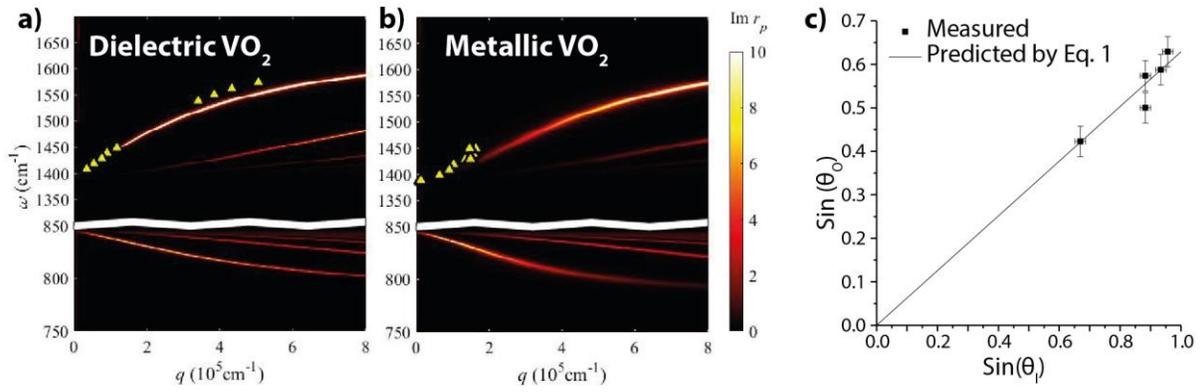

Figure 2. Hyperbolic polariton dispersion in VO$_2$ over both dielectric (a) and metallic (b) domains compared to numerical calculations. From the measured dispersion, the angle of refracted waves at 1450cm$^{-1}$ can be computed for a given incident angle and compared against experimentally measured results in (c). There has been no fitting in this result.

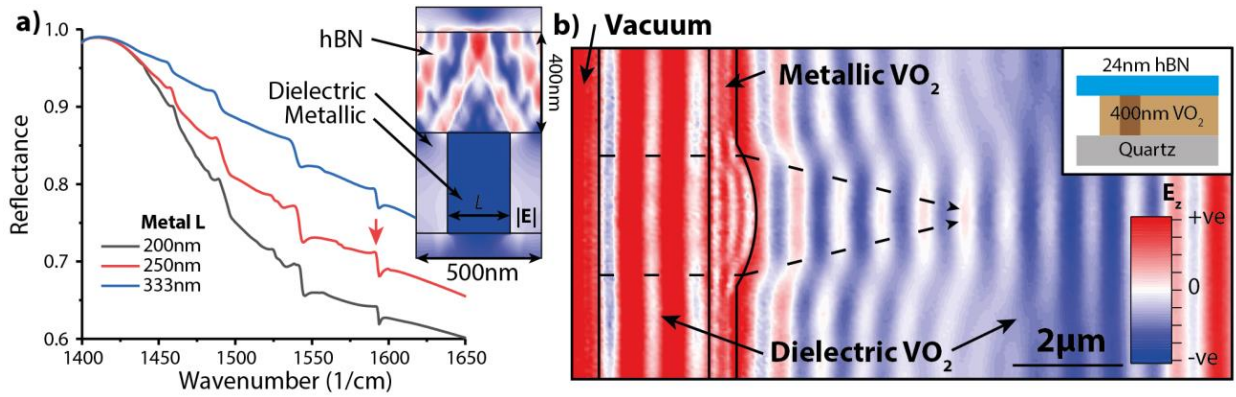

Figure 3. Schematic of refractive planar optics and reconfigurable resonators using phase-change materials (a) shows a tunable polariton resonator of hBN and VO$_2$, demonstrating a rewritable reflection profile. The pitch is 500nm, with a particle length (L). The red arrow indicates the mode which is plotted as an inset. (B) shows a simulation of a refractive polariton lens, which uses a semi-circular domain of metallic VO$_2$ to launch polariton waves at 1418cm$^{-1}$.